\documentclass[letterpaper]{article} 
\usepackage{aaai2027}  
\usepackage[hyphens]{url}  
\usepackage{graphicx} 
\usepackage{natbib}  
\usepackage{caption} 
\usepackage{algorithm}
\usepackage{algorithmic}

\usepackage{newfloat}
\usepackage{listings}
\DeclareCaptionStyle{ruled}{labelfont=normalfont,labelsep=colon,strut=off} 
\floatstyle{ruled}
\newfloat{listing}{tb}{lst}{}
\floatname{listing}{Listing}

\usepackage{booktabs}

\usepackage{makecell}
\usepackage{multirow}  
\usepackage{amsmath}
\usepackage{amssymb}
\nocopyright

\title{Pixel-TTS: Image based Text Rendering for Robust Text-to-Speech}
\author {
    Adarsh Arigala \textsuperscript{\rm 1},
    Arjun Gangwar \textsuperscript{\rm 1},
    Srinivasan Umesh\textsuperscript{\rm 1},
    Yova Kementchedjhieva \textsuperscript{\rm 2},
}
\affiliations {
    \textsuperscript{\rm 1}SPRING Lab, Indian Institute of Technology, Madras, India\\
    \textsuperscript{\rm 2}MBZUAI, UAE \\
   arigalaadarsh780@gmail.com, arjungangwar@gmail.com, umeshs@ee.iitm.ac.in, yova.kementchedjhieva@mbzuai.ac.ae
}

\begin{document}

\maketitle

\begin{abstract}
Recent advances in pixel-based text modeling show that representing text as images enables models to exploit visual cues for language understanding. Grounding text in its visual form allows structurally similar characters with different Unicode encodings to produce similar embeddings, benefiting cross-lingual and zero-shot scenarios. Conventional text-based approaches treat each character independently, limiting generalization to unseen characters and requiring embedding expansion during cross-lingual adaptation. We propose \textbf{Pixel-TTS}, a text-to-speech framework for visually grounded speech synthesis. It renders text as images and projects them through a 2D convolutional layer to generate embeddings. This design eliminates embedding matrix expansion during fine-tuning while improving robustness to unseen characters and orthographic variations. Extensive experiments show Pixel-TTS achieves competitive performance with strong baselines, faster convergence and robust zero-shot generalization.

\end{abstract}



\begin{figure*}[t]
  \centering
  \includegraphics[width=\textwidth,height=8.5cm]{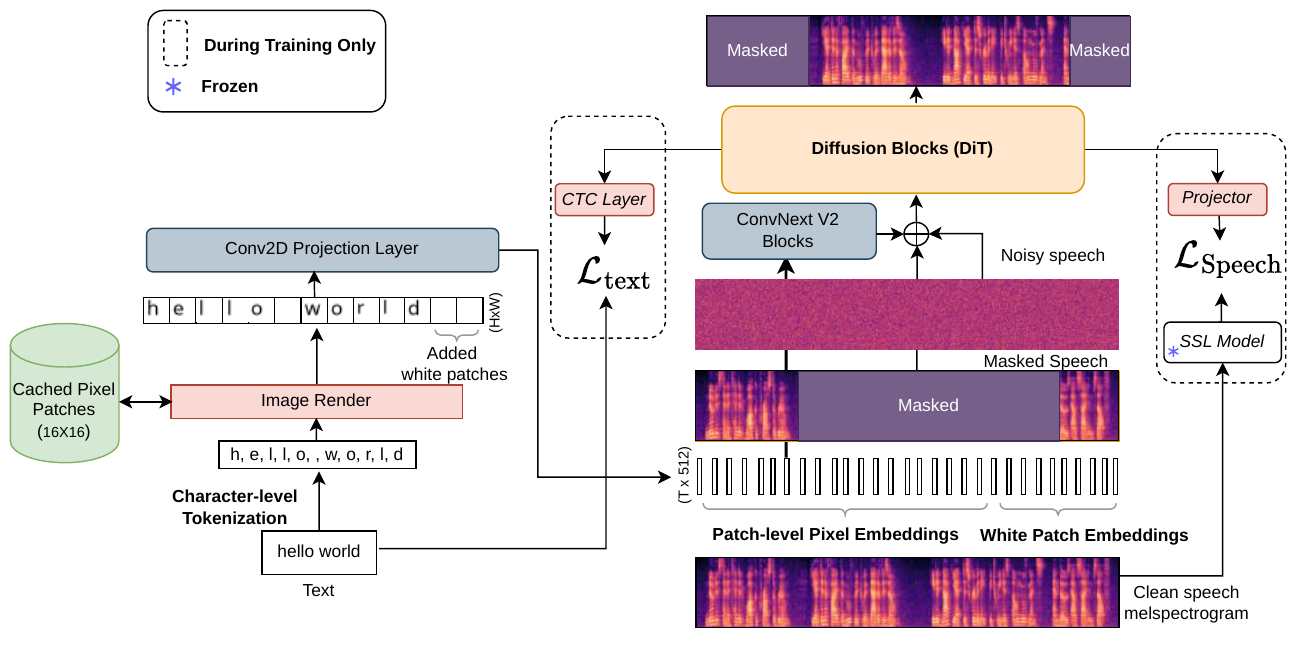}
   \caption{Overview of the proposed method.}
  \label{fig:architecture}
  
\end{figure*}

\section{Introduction}

Modern text-to-speech (TTS) systems achieve high synthesis quality but often struggle to generalize to unseen languages, particularly under zero-shot or low-resource adaptation. This limitation arises from their reliance on discrete Unicode-based embeddings, which treat each character independently and necessitate explicit vocabulary expansion during language adaptation, increasing model complexity and training cost ~\cite{eskimez2024e2ttsembarrassinglyeasy,chen-etal-2024-f5tts, choi2025accelerating}. 

Inspired by advances in machine translation, where text is rendered as images and mapped to pixel-level embeddings, pixel-based text encoding has emerged as a viable alternative to vocabulary-based methods in natural language processing (NLP), showing promising results in machine translation~\cite{salesky-etal-2021-robust,rust-etal-2023-pixel,overcoming-vocabulary-constraints}. Early studies extended visual text representations to speech generation by replacing character embeddings with rendered text images. vTTS \cite{vTTS} incorporated a lightweight CNN-based visual encoder into the conventional FastSpeech2 architecture and demonstrated that visually rendered text can synthesize speech, while also showing that visual cues such as bold, italic, and underline can control speech emphasis and emotion. In a separate line of research, \cite{VisualOnomaToWave} proposed Visual Onoma-to-Wave for environmental sound synthesis from visually rendered onomatopoeic text. Both studies relied on lightweight CNN-based visual encoders within conventional FastSpeech2-based architectures. However, the TTS framework proposed in \cite{vTTS} was evaluated only on monolingual, single-speaker datasets. Consequently, it remains unclear whether pixel-based text representations can improve robustness to unseen Unicode characters in zero-shot cross-lingual transfer and low-resource adaptation settings, exploit visual similarities among languages written in the same script, or benefit from multilingual scaling.



To address these limitations, we propose Pixel-TTS~\footnote{Source code and trained models will be released soon.}, an end-to-end TTS framework grounded in visual text representations. By rendering text as images and projecting them to pixel-level embeddings within a alignment free flow-matching TTS framework, Pixel-TTS captures perceptual and structural character similarities while improving robustness to unseen or orthographically perturbed inputs. Unlike conventional TTS systems, Pixel-TTS avoids embedding matrix expansion during cross-lingual adaptation and achieves faster convergence while maintaining natural speech synthesis quality.
In this work, we present the following contributions to text-to-speech synthesis

\begin{itemize}
    \item \textbf{End-to-end Pixel-TTS:} We introduce a visually-grounded alignment free flow-matching TTS model leveraging pixel-level embeddings, achieving faster convergence while maintaining synthesis quality comparable to competitive baselines.
    \item \textbf{Out-of-domain evaluation:} We demonstrate that Pixel-TTS generalizes effectively to unseen and accented English speech, achieving consistent improvements in intelligibility.
    \item \textbf{Improved cross-lingual generalization:} Pixel-based embeddings enable the model to transfer knowledge across languages with visually similar scripts.
    \item \textbf{Efficient fine-tuning in low-resource settings:} Experiments on low-resource German Common Voice subsets, initialized from a LibriTTS-pretrained model, show that Pixel-TTS converges faster than conventional text-based TTS models.
    \item \textbf{Robustness to orthographic noise:} Pixel-TTS handles Unicode and l33tspeak distortions better than text-based models.
    \item \textbf{Multilingual Scaling:} We train multilingual Text-TTS and Pixel-TTS models on English, German, French, and Dutch, demonstrating that Pixel-TTS better leverages shared visual patterns across related Latin scripts, resulting in improved multilingual performance.
\end{itemize}

 
\section{Methodology}
\label{methodology}
In this section, we describe the Pixel-TTS framework  (Figure.\ref{fig:architecture}). The model builds upon ADMA \cite{choi2025accelerating}, an extension of F5-TTS that leverages its dual-modality alignment to accelerate convergence. We adopt ADMA as our Text-TTS baseline, as it provides a strong benchmark with state-of-the-art performance. Pixel-TTS has three main components: (i) text-to-image rendering, (ii) projection of the rendered images, and (iii) a unified training objective combining conditional flow matching with dual-modality text and speech alignment.

\subsection{Text-to-Image}

To address text-to-speech alignment without explicit duration modeling, we adopt a character-level representation following F5-TTS~\cite{chen-etal-2024-f5tts}. Each character is rendered as a fixed $16 \times 16$ grayscale patch following the PIXEL framework~\cite{rust-etal-2023-pixel}. White $16 \times 16$ patches are used as filler tokens to preserve monotonic alignment with the audio mel-spectrogram. The number of patches is matched to the target acoustic frame length ($T$) through padding.

Rather than processing patches independently, all character patches are stacked along the width dimension to form a single image
$X \in \mathbb{R}^{H \times W},$ where $H$ denotes the patch height and $W$ is the total width of the stacked patches. In our implementation, $H=16$ and $W=T*16$, where T is the target acoustic frame length. All character patches are pre-computed and cached to improve training efficiency.

\subsection{Projection of Pixels to Embeddings}

The constructed image $X$ is projected into a sequence of embeddings using a 2D convolutional layer. The Conv2D layer is configured with input channels $=1$, output channels $=\text{d}_{\text{text}}=512$, kernel size $16 \times 16$, and stride $16$. This configuration maps each $16 \times 16$ patch into a single embedding vector, producing
$ E \in \mathbb{R}^{\text{T} \times \text{d}_{\text{text}}}, $ where $\text{T} = W / 16$ corresponds to the temporal resolution of the mel-spectrogram. This design preserves character-level temporal alignment while enabling the filters to learn visually grounded text representations aligned with speech. The resulting embeddings are processed by four stacked ConvNeXtV2 \cite{ConvNeXtV2} blocks operating at a fixed embedding dimensionality of 512, similar to text-based TTS models. This projection enables visually similar characters to produce similar embeddings, accelerating convergence (Section~\ref{visual_similar_patterns}). Apart from these component, we follow the ADMA baseline for the overall architecture and training procedure.


\begin{table*}[!t]
\centering
\small
  
\setlength{\tabcolsep}{6pt}        

\begin{tabular}{c|cccc|cccc}
\hline
 \multirow{2}{*}{\textbf{\makecell{Updates \\ (K)}} }
& \multicolumn{4}{c|}{\textbf{Text-TTS}} 
& \multicolumn{4}{c}{\textbf{Pixel-TTS}} \\

& \textbf{WER↓} & \textbf{SIM↑} & \textbf{UTMOS↑} & \textbf{CER↓}
& \textbf{WER↓} & \textbf{SIM↑} & \textbf{UTMOS↑} & \textbf{CER↓} \\
\hline

\textbf{Ground Truth} 
& \textbf{ 2.47} & \textbf{0.695} & \textbf{4.098} & \textbf{0.93} 
& \textbf{ 2.47} & \textbf{0.695} & \textbf{4.098} & \textbf{0.93}  \\

\hline

60  & 21.95 & 0.458 & 3.507 & 15.02 & 22.01 & 0.472 & 3.672 & 14.80 \\
80  & 14.78	& 0.507	& 3.748	& 10.13 & 12.53	& 0.525	& 3.867	& 8.04  \\
100 & 10.24	& 0.533	& 3.865	& 6.79  & 7.54	& 0.548	& 3.93	& 4.63 \\
120 & 7.43  & 0.548 & 3.945 & 4.71  & 5.68  & 0.558 & 3.967 & 3.40  \\
140 & 5.63	& 0.559	& 3.971	& 3.4   & 4.53	& 0.565	& 3.975	& 2.45 \\
160 & 4.68	& 0.566	& 3.973	& 2.6   & 3.56	& 0.569	& 3.985	& 1.72 \\
180 & 3.76  & 0.571 & 3.999 & 1.92  & 3.28  & 0.573 & 3.995 & 1.53 \\
200 & 3.46	& 0.575 & 4.007	& 1.65  & 2.71	& 0.579	& 4.003	& 1.18 \\
220 & 2.99	& 0.579	& 4.015	& 1.42  & 2.44	& 0.580	& 4.000	& 1.05 \\
240 & 2.84  & 0.582 & 4.020 & 1.33  & 2.53  & 0.583 & 4.018 & 0.98 \\ 
260 & 2.77	& 0.584	& 4.028	& 1.4   & 2.57	& 0.581	& 4.017	& 1.00 \\
280 & 2.51	& 0.588	& 4.041	& 1.11  & 2.16	& 0.580 & 4.011	& 0.79 \\
300 & 2.53  & 0.591 & 4.061 & 1.16  & 2.28  & 0.579 & 4.013 & 0.81 \\

\hline
\end{tabular}
\caption{Quantitative evaluation of Text-TTS and Pixel-TTS across training updates on the LibriSpeech-PC test set. Arrows indicate preferred direction: lower WER/CER (↓) and higher SIM/UTMOS (↑).}
\label{tab:training_progress}
\end{table*}


\subsection{Flow Matching and Alignment Objective}

We adopt the conditional flow matching (CFM) objective from F5-TTS~\cite{chen-etal-2024-f5tts} as the primary generative loss. CFM learns a time-dependent vector field that transports Gaussian noise to the target speech distribution through continuous-time flow integration. In Pixel-TTS, CFM is conditioned on the pixel-level embeddings obtained from the projection of the text-to-image representation, allowing the model to synthesize speech from visually grounded text features.
 Following ADMA, we incorporate auxiliary alignment losses to improve mapping between the text and speech modalities. A CTC-based text alignment loss \cite{CTC} is applied at an intermediate layer to encourage early character-level alignment. A speech representation alignment loss is introduced using HuBERT~\cite{hubert}, with features extracted from the 21\textsuperscript{st} transformer layer on 16\,kHz waveforms, and optimized via cosine similarity.   The final objective combines the CFM loss with the text and speech alignment terms:
\[
\mathcal{L} = 
\mathcal{L}_{\mathrm{CFM}}
+ \lambda_{\mathrm{text}} \mathcal{L}_{\mathrm{text}}
+ \lambda_{\mathrm{speech}} \mathcal{L}_{\mathrm{speech}},
\]
where $\lambda_{\mathrm{text}}=0.1$ and $\lambda_{\mathrm{speech}}=1.0$. This joint objective encourages faster text-speech alignment while preserving high-quality speech synthesis.

\section{Training Setup}

We follow the original ADMA implementation using the small configuration, consisting of approximately 159M parameters, 18 Transformer layers, 12 attention heads, hidden size 768. Waveform synthesis is performed using the pre-trained Vocos vocoder~\cite{siuzdak2024vocos}, trained on LibriTTS~\cite{zen2019libritts} for 300k steps. Training uses a cumulative batch size of 0.758 hours of audio. Optimization is performed with AdamW~\cite{loshchilov2019decoupled} using a learning rate of $7.5 \times 10^{-5}$, with 20k warmup steps followed by linear decay. Experiments are conducted on 8 NVIDIA A100 GPUs. \\
\noindent Audio is processed at a 24 kHz sampling rate before feature extraction. Mel-spectrograms are computed using a 1024-sample window, 256-sample hop size, and 100 mel bins. To align character-level embeddings with acoustic frames, $16 \times 16$ white patches are appended as filler tokens to character patches. Model performance is evaluated using word error rate (WER)\cite{radford2022whisper}, character error rate (CER), speaker similarity (SIM)\cite{Chen2021WavLM}, and UTMOS \cite{UTMOS} as objective measures of intelligibility, speaker preservation, and naturalness. We further conduct a mean opinion score (MOS) evaluation with human listeners to assess the subjective naturalness and overall quality of the synthesized speech.
 
\section{Dataset}

We train our model on LibriTTS~\cite{zen2019libritts}, a 585-hour multi-speaker English corpus sampled at 24\,kHz and derived from LibriSpeech~\cite{panayotov2015librispeech}.

\textbf{In-Domain Evaluation:} We follow the F5-TTS protocol using the LibriSpeech-PC~\cite{meister2023librispeechpc} test set, which contains 2.2 hours of speech comprising 1,127 utterances with durations ranging from 4--10\,s. Evaluation is performed in a cross-sentence generation setting, where prompt and reference come from same speakers.

\textbf{Out-of-Domain Evaluation:} To evaluate out-of-domain generalization, we use the SEED-TTS (English)~\cite{Anastassiou2024SeedTTSAF} and LJSpeech~\cite{ljspeech17}. The SEED-TTS (English) test set contains 1,088 evaluation samples. Since LJSpeech is a single-speaker corpus, we construct a zero-shot evaluation set from 2 hours of its training split, resulting in 1,017 samples with durations 4--10\,s.

\textbf{Evaluation on Non-Native Speakers}: To evaluate robustness on accented English, we use the L2-ARCTIC corpus~\cite{zhao2018l2arctic}, which contains speech from non-native English speakers with six first-language (L1) backgrounds (Arabic, Mandarin, Hindi, Korean,  Spanish, and Vietnamese). We use two speakers, resulting in approximately 6.52 hours of speech comprising 4,634 utterances.

\textbf{Zero-Shot Cross-Lingual Evaluation:} For zero-shot, we use the German, French, and Dutch Common Voice\cite{commonvoice-2020} test sets. The experimental setup and dataset statistics are described in Section~\ref{Zero-Shot-Cross-Lingual-evaluation}.

\textbf{Low-Resource Adaptation:} We fine-tune both Text-TTS and Pixel-TTS on 10-hour and 50-hour subsets of the German Common Voice training set. Details of the fine-tuning setup are provided in Section~\ref{finetuning}.

\textbf{Multilingual Training:} In Section~\ref{multilingual}, we extend our study to a multilingual setting. We train Pixel-TTS on a combined dataset consisting of LibriTTS (585 hrs) and German (250 hrs), French (250 hrs), and Dutch (57.18 hrs) subsets from Common Voice, resulting in a total of 1,142.18 hrs of training data. All other experiments and analyses in Sections~\ref{LibriTTS-Trained-Model} use Pixel-TTS trained only on LibriTTS.


\section{Results}
\label{LibriTTS-Trained-Model}
\subsection{Quantitative Analysis}
Table~\ref{tab:training_progress} summarizes the training progress of Text-TTS  and Pixel-TTS on the LibriSpeech-PC test set. At the same training level of 300k updates, Pixel-TTS achieves lower error rates while maintaining speaker similarity and UTMOS comparable to Text-TTS. Pixel-TTS attains a WER of 2.28 compared to 2.53 for Text-TTS, and CER of 0.81 versus 1.16, indicating improved intelligibility. Pixel-TTS leverages visual similarity between characters by exploiting shared structural patterns, leading to faster convergence. Text-TTS, however, treats each character as a separate Unicode character and requires more steps for similar performance.
As further demonstrated in Sections~\ref{Zero-Shot-Cross-Lingual-evaluation} and~\ref{finetuning}, this shared visual structure enhances generalization in cross-lingual and low-resource scenarios.

\subsection{Subjective Evaluation}

We further conducted a subjective listening evaluation to compare the perceptual quality of Text-TTS and Pixel-TTS. Sixteen participants evaluated 10 randomly selected audio samples generated by both models using a 5-point Mean Opinion Score (MOS) scale, where higher scores indicate better naturalness. Each participant independently rated both systems, resulting in a total of 160 ratings per model.

As shown in Table~\ref{tab:mos}, Pixel-TTS achieved an average MOS of \textbf{4.63 $\pm$ 0.11}, outperforming Text-TTS, which obtained a MOS of \textbf{4.25 $\pm$ 0.14}. The consistent improvement in MOS indicates that the proposed pixel-based text representation not only improves objective intelligibility metrics such as WER and CER but also enhances the perceived naturalness of synthesized speech.

\begin{table}[!h]
\centering
\setlength{\tabcolsep}{8pt}
\begin{tabular}{lc}
\hline
\textbf{Model} & \textbf{MOS} ↑ \\
\hline
Text-TTS  & 4.25 $\pm$ 0.14 \\
Pixel-TTS &  4.63 $\pm$ 0.11 \\
\hline
\end{tabular}
\caption{MOS evaluation results (mean  $\pm$ 95\% confidence interval). Higher MOS indicates better naturalness. }
\label{tab:mos}
\end{table}


 
\subsection{Out-of-Domain Evaluation}

Table~\ref{tab:out_domain} summarizes the out-of-domain evaluation results on the SEED-TTS-EN and LJ-Speech datasets. On the SEED-TTS-EN test set, Pixel-TTS achieves better intelligibility than Text-TTS, with a WER/CER of 2.19/0.90 compared to 2.36/1.06 for Text-TTS. We further evaluate the models on the LJ-Speech dataset to assess out-of-domain performance. Pixel-TTS again outperforms Text-TTS, achieving a WER/CER of 5.13/2.84 compared to 5.69/3.27. In both datasets, Pixel-TTS consistently improves intelligibility while maintaining comparable speaker similarity and perceptual speech quality.

\begin{table}[!h]
\centering

\setlength{\tabcolsep}{3pt}
\begin{tabular}{l|cccc}
\hline
\textbf{Model} & \textbf{WER} ↓  & \textbf{SIM} ↑ & \textbf{UTMOS} ↑ & \textbf{CER} ↓  \\
\hline

\multicolumn{5}{c}{\textbf{SEED-TTS-EN}} \\
\hline
Ground Truth & 1.86 & 0.734 & 3.527 & 0.80 \\

Text-TTS     & 2.36 &  0.587  &  3.926 & 1.06 \\
Pixel-TTS    &  2.19 & 0.577 & 3.890 & 0.90 \\
\hline

\multicolumn{5}{c}{\textbf{LJSpeech}} \\
\hline
Ground Truth & 3.80 & 0.768 & 4.391 & 2.93 \\
Text-TTS     & 5.69 &  0.677 & 4.275 & 3.27 \\
Pixel-TTS    &  5.13  & 0.657 & 4.151 & 2.84  \\
\hline
\end{tabular}

\caption{Out-of-domain evaluation of models trained on LibriTTS and evaluated on the SEED-TTS-EN and LJSpeech test sets. }
\label{tab:out_domain}
\end{table}

\subsection{Evaluation on Non-Native Speakers}

To further evaluate robustness to accented English speech, we evaluated the LibriTTS-trained 300k checkpoint on the L2-ARCTIC dataset. In this evaluation, the reference speech is provided by non-native English speakers, and the model is conditioned on these utterances to synthesize accented English speech. 

As shown in Table~\ref{tab:l2arctic}, Pixel-TTS achieves substantially better intelligibility than Text-TTS, obtaining a WER/CER of 3.87/1.99 compared to 5.35/3.07 for Text-TTS. While both models achieve comparable speaker similarity and perceptual speech quality, Pixel-TTS consistently produces more accurate transcriptions, demonstrating robustness to accented and non-native English speech.

\begin{table}[!h]
\centering

\setlength{\tabcolsep}{3pt}
\begin{tabular}{lcccc}
\hline
\textbf{Model} & \textbf{WER} ↓ & \textbf{SIM} ↑ & \textbf{UTMOS} ↑ & \textbf{CER} ↓ \\
\hline
Ground Truth & 11.58 & 0.758 & 3.843 & 6.07 \\ 
Text-TTS     & 5.35  &  0.585  & 4.055  & 3.07 \\
Pixel-TTS    &  3.87  & 0.584 & 4.002 &  1.99  \\
\hline
\end{tabular}

\caption{Evaluation on the L2-ARCTIC dataset containing non-native English speech.}
\label{tab:l2arctic}
\end{table}

\begin{table}[!h]
\centering
\setlength{\tabcolsep}{2pt} 
\begin{tabular}{lcccccc}
\hline
\multicolumn{6}{c}{\textit{Zero-Shot Test Set of Unseen Languages}} \\
\hline
\textbf{Lang} & \textbf{Hrs} & \textbf{\makecell{LibriTTS \\ Vocab}} & \textbf{\makecell{Unique \\ OOV}} & \textbf{\makecell{Total \\ OOV}} & \textbf{\makecell{Total \\ Chars}}  \\
\hline

German & 24.97 & 75 & 62 & 13735 & 946621  \\
French & 23.39 & 75 & 114 & 7144 & 875605   \\
Dutch  & 13.15 & 75 & 20 & 360  & 547314  \\
\hline
\multicolumn{6}{c}{\textit{Fine-tuning Datasets (German)}} \\
\hline
\textbf{Lang}  & \textbf{Hrs} & \textbf{\makecell{LibriTTS \\ Vocab}} & \textbf{\makecell{Unique \\ OOV}} & \textbf{\makecell{Total \\ OOV}} & \textbf{\makecell{Total \\ Chars}} \\  
\hline
DE-10h & 10  & 75 & 23 & 5938  & 427331   \\
DE-50h & 50  & 75 & 51 & 29760 & 2136559 \\
\hline
\end{tabular}

\caption{Statistics of Unseen Languages and Fine-tuning Datasets in TTS Evaluation.}
\label{tab:zeroshot_evaluation_durations}
\end{table}
 

\begin{table*}[!h]
\centering
\setlength{\tabcolsep}{3pt}   
\begin{tabular*}{\textwidth}{@{\extracolsep{\fill}} l | c c c c | c c c c | c c c c  }
\hline
\multirow{2}{*}{\textbf{Language}} 
& \multicolumn{4}{c|}{\textbf{Ground Truth}} 
& \multicolumn{4}{c|}{ \textbf{Text-TTS}} 
& \multicolumn{4}{c}{\textbf{Pixel-TTS}} 
 \\
 & \textbf{WER↓} & \textbf{SIM↑} & \textbf{UTMOS↑} & \textbf{CER↓} 
 &\textbf{WER↓} & \textbf{SIM↑} & \textbf{UTMOS↑} & \textbf{CER↓} 
 &\textbf{WER↓} & \textbf{SIM↑} & \textbf{UTMOS↑} & \textbf{CER↓}  \\
 
\hline
German & 6.21  & -   & 2.481 &2.18 & 71.49  &   0.481 & 3.290 & 31.50 &   66.48 & 0.481 & 3.303 & 27.36   \\
French & 12.35& -   & 2.238   &  5.39 & 63.95 & 0.411 & 3.217 & 31.71  &   62.56	& 0.400 &	 3.233 &	 29.57  \\
Dutch  & 4.67  & - &2.612 & 1.60    & 47.14 & 0.469  & 3.460 &  18.58  &  44.30	& 0.462 &  3.472 & 16.41   \\
\hline
\end{tabular*}
\caption{Zero-shot evaluation on unseen languages (300K steps). Pixel-TTS   shows better intelligibility than Text-TTS.}
\label{tab:unseen_languages}
\end{table*}

\begin{table*}[!t]
  
\centering
 
\setlength{\tabcolsep}{8pt}
\begin{tabular} {c| c|cccc|cccc}
\hline
\multirow{2}{*}{\textbf{Language}}  & \multirow{2}{*}{\textbf{\makecell{Updates \\ (K)}}} & \multicolumn{4}{c|}{\textbf{Text-TTS}} & \multicolumn{4}{c}{\textbf{Pixel-TTS}} \\
 &  & \textbf{WER↓} & \textbf{SIM↑} & \textbf{UTMOS↑} & \textbf{CER↓}  & \textbf{WER↓} & \textbf{SIM↑} & \textbf{UTMOS↑} & \textbf{CER↓}  \\
\hline
 \textbf{Ground-Truth} &  - &  \textbf{6.21}  & -   &  \textbf{2.481} & \textbf{2.18} &   \textbf{6.21}  & -   &  \textbf{2.481} & \textbf{2.18}  \\ \cline{1-10}  
\multirow{5}{*}{ de-10h  } & 10  & 125.00 & 0.361 & 2.347 & 88.70 
        &  61.02 & 0.512  & 3.268  &  24.87  \\
 & 30  & 115.79 & 0.485 & 2.520 & 83.04 
        & 38.54 &  0.582  &  3.211  &  15.61  \\
 & 50  & 100.88 & 0.558 & 2.743 & 65.26 
        & 24.03 &  0.601 &  3.123  & 9.14 \\
 & 70  & 50.63  & 0.592 & 2.943 & 27.45 
        & 16.67 & 0.601 & 3.053  & 6.42 \\
 & 90  & 27.28  & 0.595 & 2.993 & 13.47 
        & 12.66 &  0.596  &  3.008  &  4.84  \\
 & 110 & 20.45  & 0.590  &  2.994 & 9.94  
        &  10.83  & 0.589 & 2.984 &  4.20  \\
 & 130 &  18.13  &  0.581 & 2.982 & 8.81  
        & 10.02 &  0.581  & 2.965 & 3.95 \\
 & 150 &  17.22   &  0.571 & 2.983 & 8.54   
        & 9.85  &  0.571 &  2.952  & 4.00  \\
\hline
\multirow{6}{*}{  de-50h  } 
 & 10  & 125.13 & 0.360 & 2.355 & 88.74 
        &  60.85 & 0.513 & 3.273 & 24.85 \\
 & 30  & 114.07 & 0.478 & 2.515 & 82.40 
          &  38.81 & 0.583 & 3.215 & 15.47 \\
 & 50  & 97.52  & 0.561 & 2.744 & 63.02 
          &  23.96 & 0.605 & 3.114 & 9.62 \\
 & 70  & 44.46  & 0.602 & 2.943 & 23.54 
         &  16.76 & 0.609 & 3.032 & 6.70 \\
 & 90  & 23.17  & 0.609 & 2.976 & 11.25 
         &  12.70 & 0.610 & 2.990 & 5.05 \\
 & 110 &  16.92  &  0.610  &  2.972  &  8.00  
         &  10.88 & 0.609 & 2.955 & 4.26 \\

 & 130 & 14.40 & 0.610 & 2.963  &  6.82   
         &  9.79 & 0.608 & 2.933 & 3.86 \\
 & 150 & 12.87 & 0.609 & 2.966 & 5.99     
         &  9.42 & 0.606 & 2.920 & 3.62 \\

\hline
\end{tabular}
\caption{Fine-tuning results of Text-TTS versus Pixel-TTS on the Common Voice German dataset. Metrics reported are WER (\%), SIM, UTMOS, and CER (\%) at different updates for 10h and 50h subsets.}
\label{tab:finetuning_de}
\end{table*}


\subsection{Zero-Shot Cross-Lingual Evaluation}
\label{Zero-Shot-Cross-Lingual-evaluation}
To assess the generalization of Pixel-TTS beyond English, we evaluated the model on three Latin-script languages not seen during training: German, French, and Dutch. While these languages share some characters with English, they also contain new characters (e.g., diacritics, umlauts, and numerals) absent from LibriTTS. For each language, we randomly selected utterances from the Common Voice test set~\cite{commonvoice-2020} with durations of 4 to 10 seconds for speech synthesis. The detailed statistics of these test sets, including the number of unique and total out-of-vocabulary (OOV) characters, are provided in Table~\ref{tab:zeroshot_evaluation_durations}. 

The German, French, and Dutch test sets contain 62, 114, and 20 unique OOV characters, respectively. Conventional Text-TTS often treats these OOV characters as filler tokens. In our experiments, Pixel-TTS handles OOV characters seamlessly, as each character is visually rendered. Table~\ref{tab:unseen_languages} summarizes the zero-shot results at 300k training steps. SIM is not computed for the ground-truth audio, as speaker information is unavailable. Across the three languages, Pixel-TTS achieves lower error rates, demonstrating better intelligibility despite the presence of unseen Unicode characters. This improvement can be attributed to pixel-level embeddings that capture visual similarity between characters, mitigating degradation typically observed with unseen textual representations.

\subsection{Fine-tuning Results}
\label{finetuning}

To further evaluate adaptability under limited data conditions, we fine-tuned both Pixel-TTS and Text-TTS on German Common Voice training subsets of 10h and 50h using a learning rate of $7.5 \times 10^{-6}$, initializing both models from the 300k-update pretrained checkpoint. For Text-TTS, the embedding matrix was expanded to include unseen German characters and numeric symbols absent from the original LibriTTS vocabulary. Table~\ref{tab:zeroshot_evaluation_durations} shows that the 10h and 50h German Common Voice training subsets contain 23 and 51 unique new characters, respectively. These were added to the original 75-characters LibriTTS vocabulary, resulting in vocabularies of 98 and 126 characters. The embeddings for new characters were initialized as the average of existing embeddings and learned from scratch during fine-tuning\footnote{During fine-tuning, we remove the CTC alignment loss for both Text-TTS and Pixel-TTS models.}. This increases learning complexity and initially results in high WER and CER.\footnote{In zero-shot evaluation, conventional Text-TTS treats unseen characters as filler tokens.}  Pixel-TTS handles unseen symbols elegantly without any embedding expansion, enabling faster adaptation. Table~\ref{tab:finetuning_de} shows that with only 10h of adaptation data, Pixel-TTS converges substantially faster than Text-TTS. At 70k updates, Pixel-TTS achieves a lower WER/CER compared to Text-TTS at 150k updates. 
At 150k updates, Pixel-TTS further improves intelligibility and achieves 42.80\%/53.16\% relative reductions in WER/CER compared to Text-TTS. Similar trends are observed for the 50h subset, where Pixel-TTS achieves 62.30\%/71.54\% relative reductions in WER/CER at 70k updates and 26.81\%/39.57\% relative reductions at 150k updates, while maintaining comparable SIM and UTMOS scores, demonstrating faster convergence under low-resource adaptation.

\subsection{Orthographic Errors}
\label{orthographic-noises}

\begin{figure}[!h]

    \centering
    \includegraphics[width=\linewidth,height=4cm]{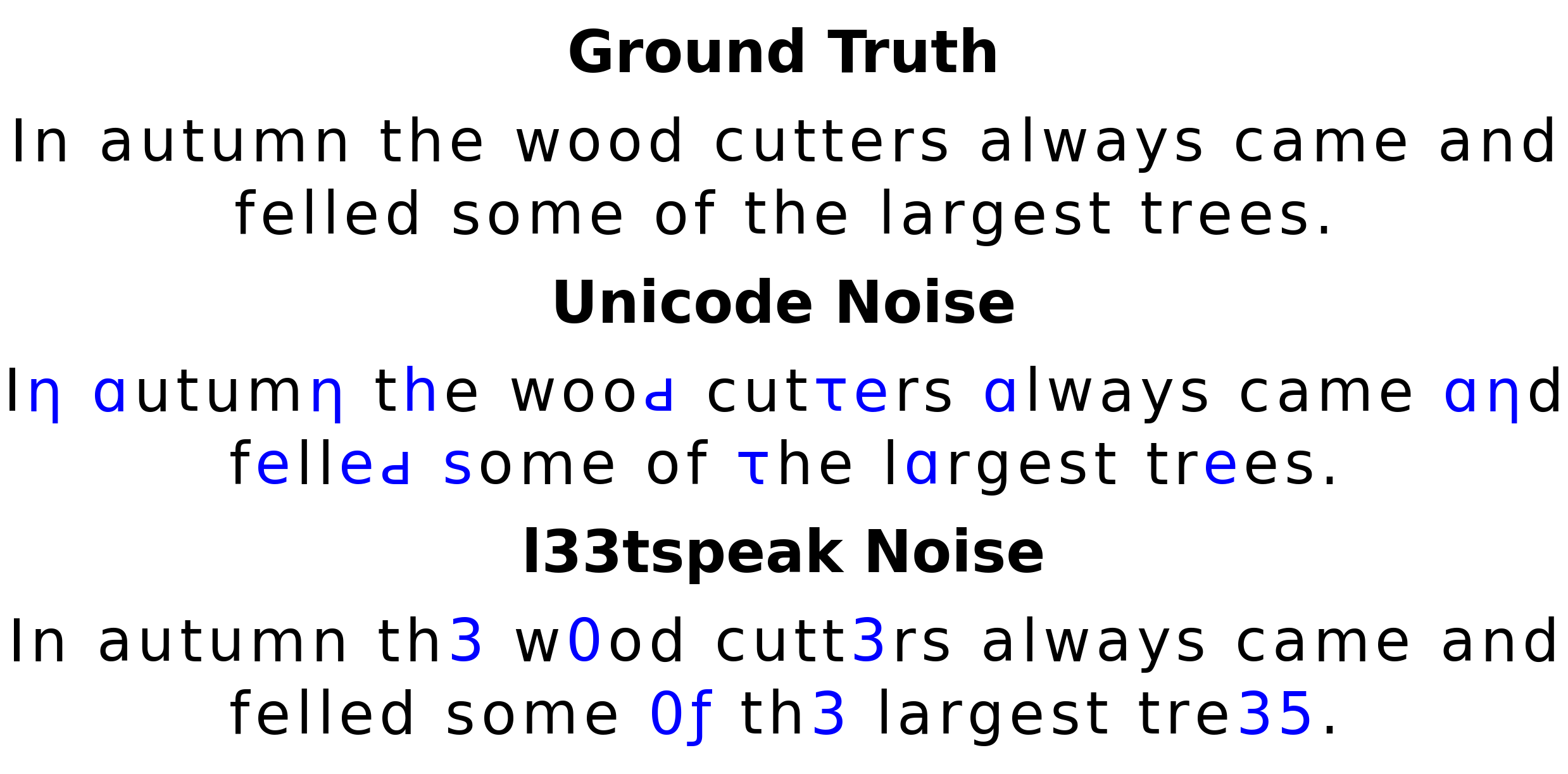}
     \captionsetup{font=small}
     \caption{Examples of synthetic text perturbations. Ground-truth sentences are shown with Unicode and l33tspeak variants; modified characters are highlighted in blue.}
     \label{fig:noise_samples}
    
\end{figure}

\begin{figure}[!h]
    \centering
  \includegraphics[width=\linewidth,height=8cm]{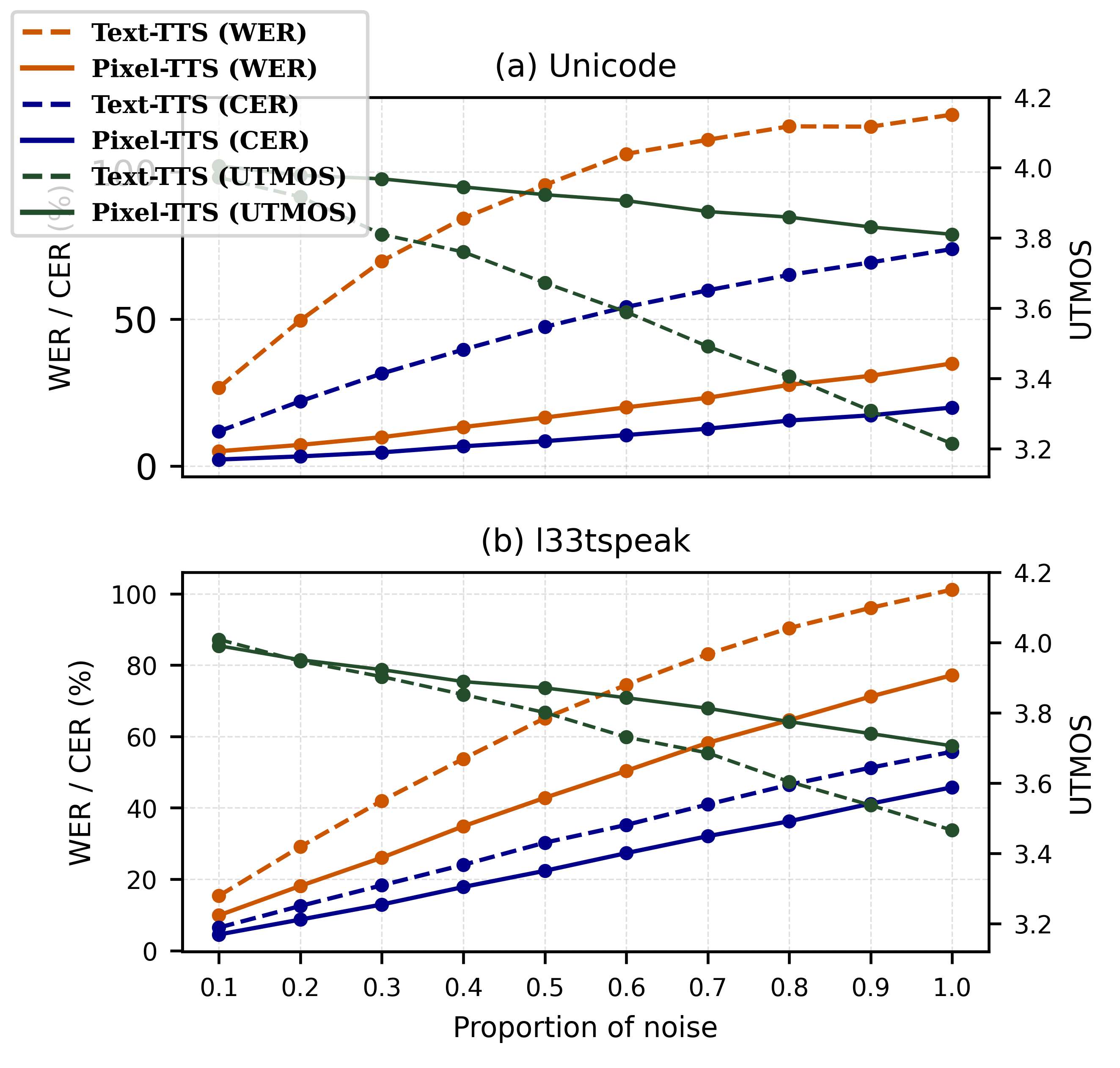}
     \captionsetup{font=small}
     \caption{Effect of synthetic character-level noise on Text-TTS and Pixel-TTS. WER/CER ↓ (left axis) and UTMOS ↑ (right axis) are shown for (a) Unicode and (b) l33tspeak perturbations.}  
     \label{fig:orthographic_mistakes}
    
\end{figure}

Inspired by the robustness experiments in~\cite{salesky-etal-2021-robust}, we evaluate Text-TTS and Pixel-TTS under character-level perturbations on the LibriSpeech-PC test set. Two types of orthographic noise are considered: Unicode homoglyph substitutions and l33tspeak transformations, applied with corruption probabilities from 0.1 to 1.0 (see Figure~\ref{fig:noise_samples} for examples and   Figure~\ref{fig:orthographic_mistakes} for quantitative results)

\textbf{Unicode Homoglyph Noise:} Standard Latin characters are replaced with visually similar Unicode variants. Text-TTS performance degrades sharply, with WER increasing from 26.67  to 119.55  (WER exceeds 100\% due to insertion errors) and UTMOS dropping from 3.972 to 3.214. Conversely, Pixel-TTS  maintains performance more robustly (WER 5.00  to 34.88 ) while relatively stable UTMOS (4.005 to 3.81), as visually similar characters produce comparable rendered patterns. 

\textbf{l33tspeak Noise:} Characters are replaced with visually similar numeric symbols. Text-TTS exhibits severe degradation (WER 15.35 to 101.15 ; UTMOS 4.009 to 3.466), whereas Pixel-TTS degrades more gradually (WER 9.87  to 77.18 ; UTMOS 3.991 to 3.706).
Across both perturbation types, Pixel-TTS demonstrates greater resilience to character-level noise by leveraging visual structure rather than relying on discrete Unicode embeddings.

\subsection{Visual Patterns in Character Embeddings}
\label{visual_similar_patterns}

\begin{figure}[!h]
    \centering
    \includegraphics[width=\linewidth ]{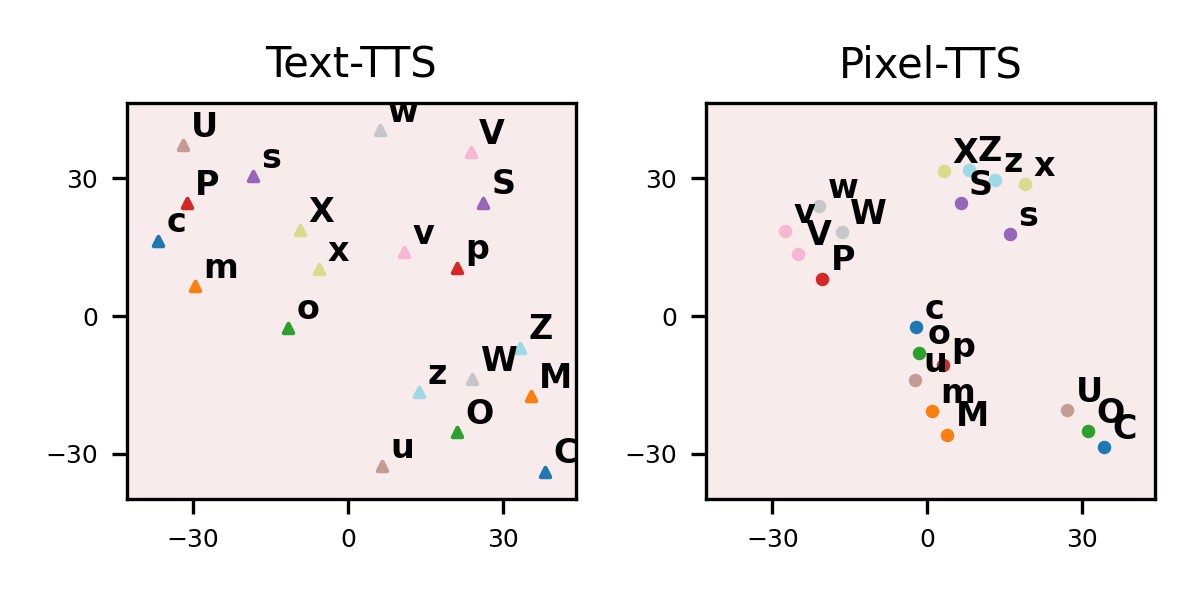}
     \captionsetup{font=small}
     \caption{t-SNE visualization of character embeddings for Pixel-TTS and Text-TTS.}
     \label{fig:tsne_visual}
    
\end{figure}

We investigate how Pixel-TTS leverages the \textbf{visual similarity} between characters by analyzing the character embeddings learned from models trained on LibriTTS. We extracted embeddings from the Text-TTS and Pixel-TTS before the ConvNeXtV2 blocks at the 60k-update checkpoint and visualized them using t-SNE, as shown in Figure~\ref{fig:tsne_visual}. The t-SNE plots reveal groupings of visually similar characters.  In conventional Text-TTS, each character is treated as a separate Unicode character, requiring many training steps to learn distinct embeddings and their relationships. Pixel-TTS exploits visual patterns, naturally clustering similar characters such as (c, C; m, M; o, O; p, P; s, S; u, U; v, V; w, W; x, X; z, Z) in the embedding space. These tighter clusters indicate that Pixel-TTS captures shared structural features across characters, facilitating faster convergence and improved generalization.

\begin{table*}[!t]
\centering
\small
\setlength{\tabcolsep}{4pt}
\begin{tabular}{l|cccc|cccc|cccc}
\hline
\multirow{2}{*}{\textbf{Dataset}} &
\multicolumn{4}{c|}{\textbf{Ground Truth}} &
\multicolumn{4}{c|}{\textbf{Text-TTS}} &
\multicolumn{4}{c}{\textbf{Pixel-TTS}} \\
 
&
\textbf{WER} ↓ &\textbf{SIM} ↑ &\textbf{UTMOS} ↑ &\textbf{CER} ↓ &
\textbf{WER} ↓ &\textbf{SIM} ↑ &\textbf{UTMOS} ↑ &\textbf{CER} ↓ &
\textbf{WER} ↓ &\textbf{SIM} ↑ &\textbf{UTMOS} ↑ &\textbf{CER} ↓ \\
\hline
LibriSpeech-PC & 2.47 & 0.695 & 4.098 & 0.93 &
2.35 & 0.619 & 4.070 & 0.83 &
 1.94 & 0.614 & 4.041 &  0.71  \\

German & 6.21 & - & 2.481 & 2.18 &
9.74 & 0.648 & 2.940 & 4.54 &
7.44 & 0.646 & 2.954 &  2.64 \\

French & 12.35 & - & 2.238 & 5.39 &
14.26 & 0.629 & 2.661 & 6.98 &
 12.14 & 0.625 & 2.684 &  5.34  \\

Dutch & 4.67 & - & 2.612 & 1.60 &
4.68 & 0.610 & 3.130 & 2.01 &
 3.82  & 0.616 & 3.154 &  1.32  \\
\hline
\end{tabular}
\caption{Evaluation of the trained multilingual Text-TTS and Pixel-TTS models on the LibriSpeech-PC, German, French, and Dutch test sets at the 160k checkpoint.}
\label{tab:multilingual_results}
\vspace{-0.3cm}
\end{table*}


\section{Multilingual Scaling}
\label{multilingual}
To investigate whether Pixel-TTS benefits from visually similar scripts under multilingual training, we construct a multilingual corpus by combining LibriTTS (585 hours) with German (250 hours), French (250 hours), and Dutch (57.18 hours) from Common Voice, resulting in approximately 1,142 hours of training data. We train both Text-TTS and Pixel-TTS using the ADMA Base architecture (335.8M parameters). The model consists of 22 DiT layers with 16 attention heads and 1024-dimensional hidden embeddings, together with a 4-layer ConvNeXtV2 text encoder blocks operating at a 512-dimensional embedding size. Training is performed for 160k updates with 45k warmup updates using an effective batch size of 4.71 hours. We evaluate the multilingual Text-TTS and Pixel-TTS models on the LibriSpeech-PC test set for English, and on the German, French, and Dutch Common Voice test sets described in Section~\ref{Zero-Shot-Cross-Lingual-evaluation}.

Table~\ref{tab:multilingual_results} summarizes the multilingual evaluation results. Across all four evaluation languages, Pixel-TTS consistently achieves lower WER and CER than Text-TTS while maintaining comparable speaker similarity (SIM) and perceptual speech quality (UTMOS). Specifically, Pixel-TTS reduces the WER/CER from 2.35/0.83 to 1.94/0.71 on English, from 9.74/4.54 to 7.44/2.64 on German, from 14.26/6.98 to 12.14/5.34 on French, and from 4.68/2.01 to 3.82/1.32 on Dutch. Significant improvements are observed on German and French, which contain many visually related characters and diacritics that are challenging for vocabulary-based representations. By learning visual patterns shared across languages, Pixel-TTS better captures similarities between characters from different languages, resulting in consistently improved intelligibility without compromising speaker similarity or perceptual speech quality. These findings demonstrate that the proposed pixel-level text representation scales effectively to multilingual training and generalizes across multiple Latin-script languages. Further analysis of the multilingual model's zero-shot cross-lingual evaluations and training progression is provided in Appendix~\ref{appendix_multilingual_zeroshot} and~\ref{appendix_multilingual_training_progress}.

\subsection{Shared Visual Representations Across Latin-Script Languages}

\begin{figure}[!h]  
    \centering
    \includegraphics[width=\linewidth ]{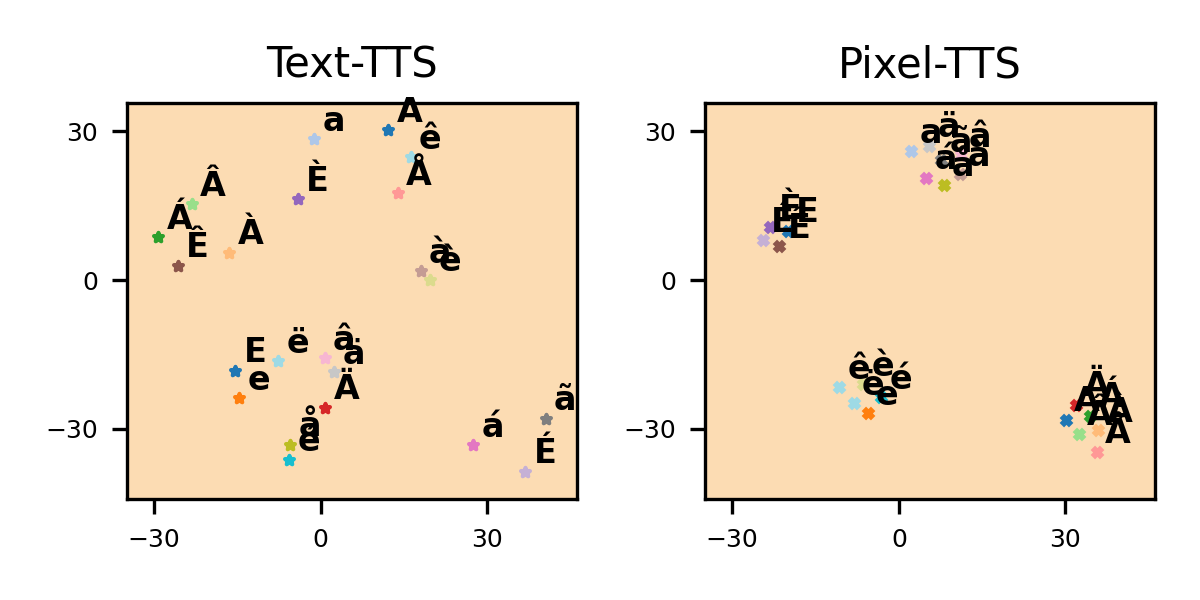}
     \captionsetup{font=small}
    \caption{t-SNE visualization of multilingual character embeddings from Pixel-TTS and Text-TTS. Pixel-TTS forms tighter clusters for visually similar Latin characters, while Text-TTS produces more dispersed representations. }
     \label{fig:tsne_multilingual}
\end{figure}

To understand how rendering text as pixels benefits multilingual TTS, we analyze the similarity between \textbf{visually related Latin characters} across multiple languages. We extract character embeddings from the multilingual Text-TTS and Pixel-TTS models before the ConvNeXtV2 blocks at the 160k-update checkpoint and compute cosine similarity between character pairs that differ only by diacritical marks, as shown in Table~\ref{tab:character_similarity}. We further visualize the learned character representations using t-SNE, as shown in Figure~\ref{fig:tsne_multilingual}.

Unlike conventional Text-TTS, where each Unicode character is represented by an independent embedding, Pixel-TTS derives representations from rendered character images and can exploit shared visual structures. For instance, character pairs such as (A–À; A–Á; a–à; and e–é) share similar visual patterns while representing distinct Unicode symbols. As shown in Table~\ref{tab:character_similarity}, Text-TTS produces low cosine similarity between these related characters, with an average similarity of \textbf{-0.002}, indicating that independently learned character embeddings do not preserve their visual relationships. Whereas, Pixel-TTS achieves a substantially higher average cosine similarity of \textbf{0.989}, demonstrating that pixel-based representations capture visual relationships among similar characters. These findings show that Pixel-TTS learns shared visual representations across Latin characters with diacritical variations, which contributes to improved generalization across languages with overlapping scripts.

 \begin{table}[!h]
\centering
\setlength{\tabcolsep}{6pt}
\begin{tabular}{lcc}
\hline
\textbf{Character Pairs} & \textbf{Text-TTS} & \textbf{Pixel-TTS} \\
\hline
$\{$\emph{A}$\} \times \{$\emph{À, Á, Â, Ä, Å}$\}$ & -0.042 & \textbf{0.991} \\
$\{$\emph{a}$\}  \times  \{$\emph{à, á, â, ã, ä}$\}$ & 0.026 & \textbf{0.984} \\
$\{$\emph{E}$\} \times \{$\emph{È, É, Ê}$\}$        & 0.019 & \textbf{0.993} \\
$\{$\emph{e}$\}  \times \{$\emph{è, é, ë}$\}$        & -0.011 & \textbf{0.988} \\
\hline
\end{tabular}
\caption{Average cosine similarity between visually similar character pairs in the learned embedding space.}
\label{tab:character_similarity}
\end{table}




 
 

     
\section{Conclusion}
 
We present Pixel-TTS, a novel text-to-speech system that encodes characters using pixel-level visual representations rather than mapping each character to conventional character embeddings used in most text-based TTS systems. By leveraging visual similarity between characters, Pixel-TTS efficiently handles unseen symbols and orthographic noise without expanding the embedding matrix during fine-tuning. Experiments across in-domain, out-of-domain, and zero-shot cross-lingual settings demonstrate that Pixel-TTS achieves lower WER and CER while maintaining comparable SIM and UTMOS scores, with faster convergence and strong generalization. Fine-tuning on German Common Voice subsets further confirms rapid adaptation to new characters. Studies on character-level perturbations demonstrate robustness, with pixel-level embeddings outperforming traditional text representations. Furthermore, multilingual analysis shows that Pixel-TTS learns shared visual representations across Latin-script languages, enabling visually similar characters to benefit from shared character structures and improving multilingual generalization.




\bibliography{aaai2027}

\appendix

\section*{Appendix}
\label{appendix_multilingual}
 
\section{Multilingual Model : Zero-Shot  Cross-Lingual  Evaluation on Unseen Languages}
\label{appendix_multilingual_zeroshot}
To further evaluate the generalization capability of the trained multilingual model, we perform zero-shot speech synthesis on three unseen Latin-script languages: Spanish, Italian, and Portuguese. Following the zero-shot evaluation protocol described in Section 5.5, we construct evaluation sets from the Common Voice test split by selecting utterance pairs with 4 to 7\,s reference speech and target text corresponding to 4 to 10\,s of speech. Table~\ref{tab:multilingual_zeroshot_dataset_stats} summarizes the statistics of the resulting evaluation sets, including the number of out-of-vocabulary (OOV) characters with respect to the multilingual training vocabulary, while Table~\ref{tab:zeroshot_unseen_languages} reports the zero-shot performance of Multilingual Text-TTS and Pixel-TTS at the 160k checkpoint.

\begin{table}[!h]
\centering
\small
\setlength{\tabcolsep}{3pt}
\begin{tabular}{lcccc}
\hline
\textbf{Language} &
\textbf{Total Hours} &
\textbf{\makecell{Number of \\ Utterances}} \\

\hline
Spanish    & 25.08 & 14290    \\
Italian    & 24.49 & 13510   \\
Portuguese & 9.44  & 5913   \\
\hline
\end{tabular}
\caption{Statistics of the zero-shot evaluation sets constructed from the Common Voice test split for unseen Latin-script languages.}
\label{tab:multilingual_zeroshot_dataset_stats}
\end{table}

\begin{table*}[!t]
\centering
\small
\setlength{\tabcolsep}{4pt}
\begin{tabular}{l|cccc|cccc|cccc}
\hline
\multirow{2}{*}{\textbf{Language}} &
\multicolumn{4}{c|}{\textbf{Ground Truth}} &
\multicolumn{4}{c|}{\textbf{Text-TTS}} &
\multicolumn{4}{c}{\textbf{Pixel-TTS}} \\
\cline{2-13}
&
\textbf{WER} ↓ & \textbf{SIM} ↑ & \textbf{UTMOS} ↑ & \textbf{CER} ↓ &
\textbf{WER} ↓ & \textbf{SIM} ↑ & \textbf{UTMOS} ↑ &\textbf{CER} ↓ &
\textbf{WER} ↓ &\textbf{SIM} ↑ &\textbf{UTMOS} ↑ &\textbf{CER} ↓ \\
\hline
Spanish    & 7.54  & --    & 2.201 & 2.53 & 24.32 & 0.492 & 2.612 & 10.19 &  22.61  & 0.479 & 2.636 &  7.89 \\
Italian    & 8.41  & --    & 2.285 & 2.49 & 31.06 & 0.539 & 2.654 & 12.86 &  26.62  & 0.518 & 2.675 &  8.88 \\
Portuguese & 9.85  & --    & 2.452 & 3.21 & 40.04 & 0.486 & 2.761 & 18.07 &  39.86  & 0.475 & 2.794 & 15.69 \\
 
\hline
\end{tabular}
\caption{Zero-shot evaluation of the multilingual Text-TTS and Pixel-TTS models on unseen Latin-script languages at the 160k checkpoint. Pixel-TTS consistently achieves lower WER and CER than Text-TTS on Spanish, Italian, and Portuguese while maintaining comparable speaker similarity (SIM) and naturalness (UTMOS).}
\label{tab:zeroshot_unseen_languages}
\end{table*}

\section{Quantitative evaluation of the multilingual model across training
updates. }
\label{appendix_multilingual_training_progress}
Tables~\ref{tab:training_progression}, \ref{tab:german_training_progression}, \ref{tab:french_training_progression}, and \ref{tab:dutch_training_progression} present checkpoint-wise evaluations of the multilingual Text-TTS and Pixel-TTS models on the English (LibriSpeech-PC), German, French, and Dutch test sets, respectively. Across all four evaluation sets, Pixel-TTS consistently converges faster, surpasses Text-TTS in terms of WER and CER, and maintains comparable speaker similarity (SIM) and naturalness (UTMOS).

 \begin{table*}[h]
\centering
\small
\setlength{\tabcolsep}{4pt}
\begin{tabular}{c|cccc|cccc}
\hline
\multirow{2}{*}{\textbf{\makecell{Updates \\ (k)}}} &
\multicolumn{4}{c|}{\textbf{Text-TTS}} &
\multicolumn{4}{c}{\textbf{Pixel-TTS}} \\
\cline{2-9}
& \textbf{WER} ↓ & \textbf{SIM} ↑  & \textbf{UTMOS} ↑  & \textbf{CER} ↓
& \textbf{WER} ↓ & \textbf{SIM} ↑  & \textbf{UTMOS} ↑  & \textbf{CER} ↓ \\
\hline
Ground Truth & 2.47 & 0.695 & 4.098 & 0.93 & 2.47 & 0.695 & 4.098 & 0.93 \\
60  & 9.04  & 0.520 & 3.799 & 5.56 & 13.45 & 0.500 & 3.589 & 8.45 \\
80  & 4.40  & 0.577 & 3.917 & 2.21 & 5.25  & 0.567 & 3.841 & 3.03 \\
100 & 2.98  & 0.599 & 4.004 & 1.20 & 2.87  & 0.593 & 3.950 & 1.32 \\
120 & 2.64  & 0.612 & 4.037 & 1.05 & 2.47  & 0.604 & 3.995 & 1.18 \\
140 & 2.43  & 0.615 & 4.063 & 0.90 &  1.85 & 0.613 & 4.016 & 0.67 \\
160 & 2.35 &  0.619 &  4.070  &  0.83 &
1.94 &  0.614  &  4.041  & 0.71 \\
\hline
\end{tabular}
\caption{Quantitative evaluation of the multilingual Text-TTS and Pixel-TTS models on the on LibriSpeech-PC test set across training updates. . Pixel-TTS consistently achieves lower WER and CER than Text-TTS while maintaining comparable speaker similarity (SIM) and naturalness (UTMOS).}
\label{tab:training_progression}
\end{table*}

\begin{table*}[h]
\centering
\small
\setlength{\tabcolsep}{4pt}
\begin{tabular}{c|cccc|cccc}
\hline
\multirow{2}{*}{\textbf{\makecell{Updates \\ (k)}}} &
\multicolumn{4}{c|}{\textbf{Text-TTS}} &
\multicolumn{4}{c}{\textbf{Pixel-TTS}} \\
\cline{2-9}
& \textbf{WER} ↓ & \textbf{SIM} ↑  & \textbf{UTMOS} ↑  & \textbf{CER} ↓
& \textbf{WER} ↓ & \textbf{SIM} ↑  & \textbf{UTMOS} ↑  & \textbf{CER} ↓ \\
\hline
Ground Truth & 6.21 & - & 2.481 & 2.18 & 6.21 & - & 2.481 & 2.18 \\
60  & 27.55 & 0.529 & 2.762 & 15.44 & 29.70 & 0.540 & 2.654 & 14.80 \\
80  & 17.77 & 0.586 & 2.846 & 9.16 & 14.75 & 0.591 & 2.810 & 6.51 \\
100 & 14.33 & 0.614 & 2.917 & 7.27 & 10.17 & 0.614 & 2.907 & 4.01 \\
120 & 11.44 & 0.631 & 2.947 & 5.37 & 8.47 & 0.629 & 2.939 & 3.17 \\
140 & 10.68 & 0.641 & 2.955 & 5.08 & 7.65 & 0.639 & 2.955 & 2.72 \\
160 & 9.74 &  0.648  & 2.940 & 4.54 & 7.44 &  0.646 & 2.954 &  2.64 \\
\hline
\end{tabular}
\caption{Quantitative evaluation of the multilingual Text-TTS and Pixel-TTS models on the German test set across training updates. Pixel-TTS consistently achieves lower WER and CER than Text-TTS while maintaining comparable speaker similarity (SIM) and naturalness (UTMOS).}
\label{tab:german_training_progression}
\end{table*}

\begin{table*}[h]
\centering
\small
\setlength{\tabcolsep}{4pt}
\begin{tabular}{c|cccc|cccc}
\hline
\multirow{2}{*}{\textbf{\makecell{Updates \\ (k)}}} &
\multicolumn{4}{c|}{\textbf{Text-TTS}} &
\multicolumn{4}{c}{\textbf{Pixel-TTS}} \\
\cline{2-9}
& \textbf{WER} ↓ & \textbf{SIM} ↑  & \textbf{UTMOS} ↑  & \textbf{CER} ↓
& \textbf{WER} ↓ & \textbf{SIM} ↑  & \textbf{UTMOS} ↑  & \textbf{CER} ↓ \\
\hline
Ground Truth & 12.35 & - & 2.238 & 5.39 & 12.35 & - & 2.238 & 5.39 \\
60  & 29.94 & 0.530 & 2.538 & 17.26 & 34.65 & 0.532 & 2.410 & 18.79 \\
80  & 21.18 & 0.579 & 2.597 & 11.41 & 19.82 & 0.577 & 2.557 & 9.76 \\
100 & 17.77 & 0.602 & 2.651 & 9.30 & 15.29 & 0.597 & 2.640 & 7.10 \\
120 & 15.50 & 0.616 & 2.673 & 7.76 & 13.61 & 0.612 & 2.670 & 6.26 \\
140 & 14.98 & 0.624 & 2.675 & 7.43 & 12.68 & 0.619 & 2.682 & 5.65 \\
160 & 14.26 &  0.629 & 2.661 & 6.98 & 12.14 & 0.625  &  2.684 &  5.34 \\
\hline
\end{tabular}
\caption{Quantitative evaluation of the multilingual Text-TTS and Pixel-TTS models on the French test set across training updates. Pixel-TTS consistently improves WER and CER over Text-TTS while maintaining comparable speaker similarity (SIM) and naturalness (UTMOS).}
\label{tab:french_training_progression}
\end{table*}

\begin{table*}[!t]
\centering
\small
\setlength{\tabcolsep}{4pt}
\begin{tabular}{c|cccc|cccc}
\hline
\multirow{2}{*}{\textbf{\makecell{Updates \\ (k)}}} &
\multicolumn{4}{c|}{\textbf{Text-TTS}} &
\multicolumn{4}{c}{\textbf{Pixel-TTS}} \\
\cline{2-9}
& \textbf{WER} ↓ & \textbf{SIM} ↑  & \textbf{UTMOS} ↑  & \textbf{CER} ↓
& \textbf{WER} ↓ & \textbf{SIM} ↑  & \textbf{UTMOS} ↑  & \textbf{CER} ↓ \\
\hline
Ground Truth & 4.67 & - & 2.612 & 1.60 & 4.67 & - & 2.612 & 1.60 \\
60  & 19.33 & 0.528 & 2.965 & 10.04 & 24.41 & 0.519 & 2.887 & 11.94 \\
80  & 11.08 & 0.571 & 3.046 & 5.38 & 10.06 & 0.575 & 3.035 & 4.28 \\
100 & 7.53 & 0.592 & 3.115 & 3.63 & 5.96 & 0.595 & 3.124 & 2.30 \\
120 & 5.90 & 0.601 & 3.143 & 2.67 & 4.81 & 0.606 & 3.145 & 1.74 \\
140 & 5.12 & 0.607 & 3.143  & 2.33 & 4.13 & 0.614 &  3.156  & 1.40 \\
160 & 4.68 & 0.610 & 3.130 & 2.01 & 3.82 & 0.616 & 3.154 &1.32 \\
\hline
\end{tabular}
\caption{Quantitative evaluation of the multilingual Text-TTS and Pixel-TTS models on the Dutch test set across training updates. Pixel-TTS consistently achieves lower WER and CER than Text-TTS while maintaining comparable speaker similarity (SIM) and naturalness (UTMOS).}
\label{tab:dutch_training_progression}
\end{table*}


\end{document}